\documentclass[sigconf]{acmart}
\usepackage{multirow}
\usepackage{colortbl}
\usepackage{makecell}
\usepackage[inline]{enumitem}

\theoremstyle{definition}

\acmConference[RecSys '25]{}{September 22--26,
  2018}{Prague, Czech Republic}

\AtBeginDocument{%
  }

\copyrightyear{2025}
\acmYear{2025}
\setcopyright{rightsretained}
\acmConference[RecSys '25]{Proceedings of the Nineteenth ACM Conference on Recommender Systems}{September 22--26, 2025}{Prague, Czech Republic}
\acmBooktitle{Proceedings of the Nineteenth ACM Conference on Recommender Systems (RecSys '25), September 22--26, 2025, Prague, Czech Republic}\acmDOI{10.1145/3705328.3759303}
\acmISBN{979-8-4007-1364-4/2025/09}

\begin{document}

\title{Describe What You See with Multimodal Large Language Models to Enhance Video Recommendations}


\author{Marco De Nadai}
\orcid{0000-0001-8466-3933}
\email{mdenadai@spotify.com}
\affiliation{%
   \institution{Spotify}
   \country{Denmark}
 }

 \author{Andreas Damianou}
\orcid{0009-0007-7194-4155}
 \email{andreasd@spotify.com}
 \affiliation{%
   \institution{Spotify}
   \country{United Kingdom}
 }
 \author{Mounia Lalmas}
\orcid{0000-0002-3531-3096}
 \email{mounial@spotify.com}
 \affiliation{%
   \institution{Spotify}
   \country{United Kingdom}
 }

\begin{abstract}
Existing video recommender systems rely primarily on user-defined metadata or on low-level visual and acoustic signals extracted by specialised encoders. These low-level features describe what appears on the screen but miss deeper semantics such as intent, humour, and world knowledge that make clips \emph{resonate with viewers}. For example, is a 30‑second clip simply a singer on a rooftop, or an ironic parody filmed amid the fairy chimneys of Cappadocia, Turkey? Such distinctions are critical to personalised recommendations yet remain invisible to traditional encoding pipelines.
In this paper, we introduce a simple, recommendation system‑agnostic zero-finetuning framework that injects high‑level semantics into the recommendation pipeline by prompting an off‑the‑shelf Multimodal Large Language Model (MLLM) to summarise each clip into a rich natural‑language description (e.g. ``a superhero parody with slapstick fights and orchestral stabs''), bridging the gap between raw content and user intent. We use MLLM output with a state‑of‑the‑art text encoder and feed it into standard collaborative, content‑based, and generative recommenders. 
On the MicroLens‑100K dataset, which emulates user interactions with TikTok‑style videos, our framework consistently surpasses conventional video, audio, and metadata features in five representative models. Our findings highlight the promise of leveraging MLLMs as on‑the‑fly knowledge extractors to build more intent‑aware video recommenders.

\end{abstract}







\maketitle

\begin{figure*}[!ht]
\label{img:teaser}
\includegraphics[width=0.9\linewidth]{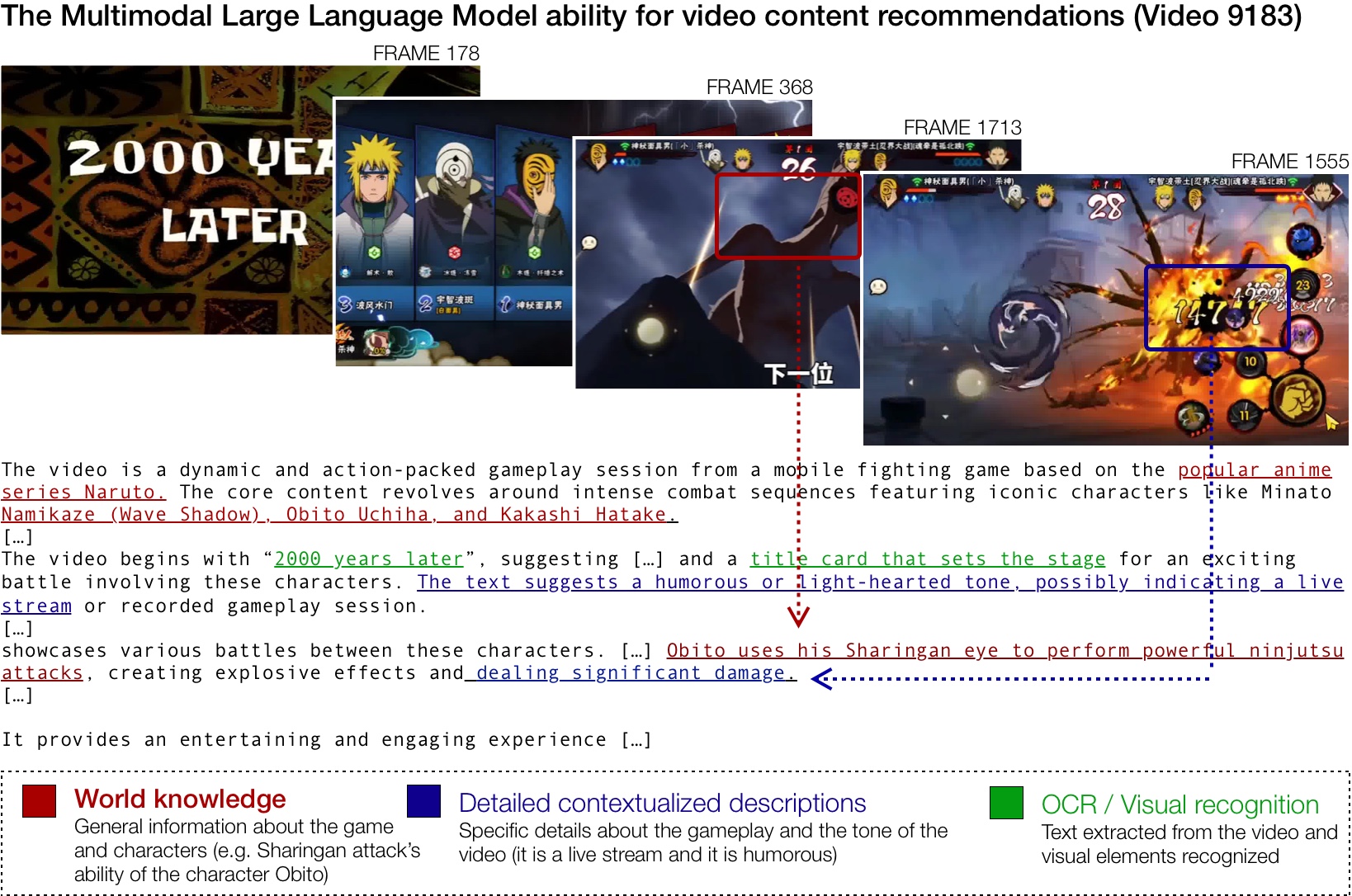}
\caption{A qualitative breakdown of a gameplay clip from a Naruto mobile fighting game (video ID 9183), demonstrating the potential of MLLMs. A MLLM (here, Qwen-VL) (i) extracts on-screen text via OCR (green), (ii) grounds entities and actions in the franchise's world knowledge (red), and (iii) composes fine-grained, time-aligned contextual descriptions of the battle dynamics (blue). These detailed outputs could resonate with users, thereby improving the recommendation system's performance.}
\Description{A qualitative breakdown of a gameplay clip from a Naruto mobile fighting game (video ID 9183), demonstrating the potential of MLLMs. A MLLM (here, Qwen-VL) (i) extracts on-screen text via OCR (green), (ii) grounds entities and actions in the franchise's world knowledge (red), and (iii) composes fine-grained, time-aligned contextual descriptions of the battle dynamics (blue). These detailed outputs could resonate with users, thereby improving the recommendation system's performance.}
\label{img:teaser}
\end{figure*}
\section{Introduction}

The rapid rise of video platforms such as TikTok, Instagram Reels, and YouTube Shorts has highlighted the critical importance of developing effective video recommendation systems~\cite{zheng2022dvr,yu2023improving,gong2022real,yuan2022tenrec}.
In fact, TikTok's success is often attributed in part to its recommendation algorithm, which is estimated to align with user interests 30\% to 50\% of the time, alongside other factors such as upload timing and social network effects~\cite{vombatkere2024tiktok}. This surge in popularity has not only underscored the influence of recommendation algorithms, but also exposed the limitations of conventional approaches in understanding what truly drives user engagement.

Indeed, conventional content‑based models distill each clip into low‑level visual, audio, and textual embeddings derived from pre-trained encoders or sparse metadata~\cite{zheng2022dvr,yu2023improving,gong2022real,yuan2022tenrec}. Although these representations capture motion, colour, or keywords, they remain blind to higher-order semantics and cultural references; the very cues that make a clip \emph{resonate with viewers}.
For instance, optical flow can reveal that "someone is dancing on a rooftop" but not that the dance parodies a 1990s superhero trope.
This lack of semantic understanding limits the ability of traditional systems to fully capture the \textit{why} behind user preferences, particularly for content rich in implicit meaning or tied to world knowledge.

In this paper, we show that Multimodal Large Language Models (MLLMs)~\cite{wang2024qwen2, chu2023qwen, ataallah2024minigpt4, liu2023llava, liu2023improvedllava} constitute a promising solution.  MLLMs can jointly analyze video frames, audio signals, and metadata to generate semantically rich captions that describe video content in detail. These captions encapsulate not only observable content, but also intent, style, and context, serving as effective proxies for user preferences. 
For example, given a gameplay footage from an anime-style fighting game (e.g. Naruto), a MLLM could recognize the characters’ signature lines and link them back to the series’ world, thereby placing the clip in a context that fans will recognise (see Figure~\ref{img:teaser}).

In more detail, this paper introduces a recommendation system \emph{model‑agnostic} framework that plugs MLLM‑generated data into existing recommender pipelines without any major change. We run a zero‑shot prompting recipe on open weights MLLMs to describe each clip, concatenate video and audio captions, and feed the resulting text to standard recommenders. Using MLLMs to generate multimodal captions, our approach bridges the gap between raw content and user intent, enabling systems to better understand and align with user preferences.

To evaluate the effectiveness of this framework, we conduct extensive experiments on the publicly available MicroLens-100K dataset~\cite{ni2023content}, a large-scale video dataset collected from a real-world video platform. We compare MLLM-generated features against traditional video, audio, and metadata-based features across various recommendation architectures, including embedding-based and generative models. 
Our results demonstrate that MLLM-derived features significantly outperform traditional baselines. Notably, multimodal captions, generated by analyzing both video and audio inputs jointly, yield the most robust representations, leading to substantial improvements in recommendation quality.

The contributions of this paper are summarized as follows:
\begin{itemize}
  \item We present the first systematic and reproducible study of leveraging open-weight MLLMs for video recommendation. This allows practitioners to use publicly available models in both academic and industry settings.
  \item We release a lightweight, \emph{zero‑finetuning} pipeline that converts clips into rich textual representations. This zero-finetuning approach is particularly advantageous as end-to-end adaptation of the requisite large-scale MLLMs is often infeasible due to substantial computational and temporal costs.
  \item Through extensive experiments on the MicroLens-100K dataset, we show that MLLM captions consistently outperform visual, audio and metadata baselines. These gains are more pronounced in videos longer than 30 seconds.
  \item We make our prompts and generated data publicly available to foster reproducibility.\footnote{\url{https://huggingface.co/datasets/marcodena/video-recs-describe-what-you-see}}
  \item Our findings underscore the transformative potential of MLLMs in video recommendation tasks, offering a pathway to systems that deliver more accurate, contextually rich, and user-aligned recommendations.
\end{itemize}

\section{Related Work}

\textbf{Video Recommendations}. The lack of publicly available large-scale video recommendation datasets with raw videos has made it difficult to develop and research novel recommender system models~\cite{ni2023content}. As a result, most published research relies on collaborative signals (e.g. \cite{gu2024modeling}), metadata, pre-computed features computed from images (e.g. \cite{yu2022improving, zheng2022dvr, lei2021semi, li2024multi, zhong2024predicting, li2024multi}) or video features extracted by proprietary systems (e.g. \cite{lee2017large}), where raw data is kept hidden~\cite{abu2016youtube}. For example, MMGCN~\cite{wei2019mmgcn} use both image covers and textual metadata in a Graph Neural Network to leverage multi-modal information. Lee \emph{et al.}~\cite{lee2017large} propose a content-based similarity learning recommendation system on 8 million videos. Singh \emph{et al.}~\cite{singh2024better} use video embeddings for generative recommendations at Youtube. 

Fortunately, the release of MicroLens~\cite{ni2023content} has allowed the direct use of raw videos to extract multiple types of information (e.g. video embeddings with custom pipelines, images) in academic research. For example, Jiang \emph{et al.}~\cite{jiang2020aspect} leverage video features to identify user interest groups for recommendation purposes, while Ni \emph{et al.}~\cite{ni2023content} finetune video encoders specifically for recommendation systems. In contrast to these approaches, our method adopts a \emph{zero-finetuning}, \emph{model-agnostic} framework that demonstrates how MLLMs can improve video recommendation performance.

\textbf{MLLMs}. Building on the long-standing AI goal of vision-language integration, MLLMs have surged in popularity by leveraging the power of LLMs for unprecedented cross-modal understanding and generation~\cite{wang2024qwen2, chu2023qwen, ataallah2024minigpt4, liu2023llava, liu2023improvedllava}. These models can process and synthesize information from diverse inputs, such as images, video frames, and audio signals, to produce rich, contextually aware textual descriptions. This capability allows MLLMs to move beyond superficial feature extraction, enabling them to identify complex entities, understand actions, and ground observations in extensive world knowledge, as demonstrated in tasks such as detailed scene description and event summarization.

For this reason, adaptations of MLLMs for recommendations are starting to emerge. For example, Peixuan \emph{et al.}~\cite{qi2024movie} trains an MLLM from scratch to classify movies into genres, suggesting that they can be used for recommendations. Fu \emph{et al.}~\cite{fu2024efficient} finetune MLLMs for sequential recommendations. Zhou \emph{et al.}~\cite{10.1145/3696410.3714764} benchmarks various image MLLMs to assess whether they can be used directly for Amazon item recommendations. Ye~\emph{et al}~\cite{ye2025harnessing} describe user preferences through text and finetune an MLLM on user interactions to predict user interests from textual metadata and the cover image of videos. 

While MLLMs are beginning to be explored in recommendation contexts, a significant gap remains. To the best of our knowledge, no prior work has systematically examined how open-weight MLLMs can be effectively leveraged to capture the complex interplay of world knowledge, aesthetic features, intent, style, and cultural references embedded in video content, which are factors essential for truly understanding and aligning with user preferences in video recommendation systems. Notably, our approach does not require finetuning and does not generate recommendations directly, offering a more scalable and efficient pathway for video recommendations. 

Our work directly addresses this gap by proposing a model-agnostic framework that extracts these rich, high-level semantic features from videos using MLLMs, demonstrating their superior performance over traditional methods.

\section{Research Hypotheses}  
We investigate whether MLLMs can systematically improve the quality of video recommendations. Our core research question is: \emph{Do MLLM-derived captions outperform classical content features in standard ranking tasks?}

MLLMs capture nuanced intent and contextual information often missed by traditional features. We hypothesize that simply replacing traditional visual, audio, and sparse textual features with semantically rich MLLM-generated descriptions will result in higher HR@K and nDCG@K scores on MicroLens-100K.

To test this, we train two representative models: one using classical video, metadata, and audio features, and another using MLLM-generated captions from videos and audio content. We evaluate both models on MicroLens-100K using a global time-based split to mimic a production setting, across multiple days.

\section{The Framework}  
Our framework addresses the semantic sparsity inherent in traditional video recommendation systems through four guiding principles. {\bf First}, we use widely available, state-of-the-art open-weight models. {\bf Second}, we focus on cross-modal grounding to reduce ambiguities by fusing audio, video and textual signals. {\bf Third}, we use frozen MLLMs, which lowers computational costs and enables frequent, low-cost training of recommendation models. {\bf Finally}, our framework is designed to be backbone-agnostic: MLLM-generated embeddings can be seamlessly integrated into any recommendation architecture (e.g., two-towers models, generative models).

\section{Experimental Results}
\label{sec:experiments}

\subsection{Setting}

\subsubsection{Data}
We use the publicly available MicroLens-100K~\cite{ni2023content}, a large-scale content-driven video dataset sourced from a real-world video mobile platform with approximately 34 million users. On this platform, creators upload and share vertical format videos ranging from a few seconds to around 10 minutes in length. 
MicroLens-100K includes 100,000 users, 19,738 items, and 719,405 interactions, resulting in a sparsity of 99.96\%. Following Ni \emph{et al.}~\cite{ni2023content}, we retain users with at least two interactions and limit each user's interaction history to their ten most recent interactions.
The dataset contains raw, full-length video files along with associated metadata, making it particularly suitable for evaluating our framework.
While our framework is capable of processing longer videos (up to several hours), we focus on short-term content for computational efficiency. Specifically, we restrict videos to a maximum length of 4 minutes and downscale them to a resolution of $426\times224$ pixels.

\subsubsection{Modalities}
We consider various state-of-the-art models and modalities to represent videos. To ensure a fair comparison, we select encoders with comparable parameter counts.

\textit{Metadata} uses a sentence text encoder that generate embeddings from video titles (written by the video creators). We use BGE-large~\cite{bge_embedding}, a 335M parameters text encoder built on top of~BERT.

\textit{Audio and Video} use dedicated audio and video encoders, respectively, to encode the content of the video. For audio, we use CLAP~\cite{wu2023large}, a 194M parameters model trained on a diverse set of (audio, text) pairs. CLAP is not speech-specific, making it well-suited for videos that include both music and speech. Video features are extracted using VideoMAE~\cite{tong2022videomae}, following Ni \emph{et al.}~\cite{ni2023content}. VideoMAE is a 307M parameters model that processes only 16 frames per video. To best capture the temporal dynamics, we extract these frames from the first 30 seconds. In line with recent findings, we observe that averaging multiple VideoMAE embeddings degrades performance~\cite{tong2022videomae}. We also tested the recently released Google VideoPrism~\cite{zhao2024videoprism},  available on Hugging Face as of June 2025, but found no performance improvement on our dataset.

\textit{MLLM (audio, video)} use \text{Qwen-VL} \cite{wang2024qwen2}, an open-weights MLLM pre-trained on 1.4B image-text pairs and videos, chosen for its ability to generate knowledge-grounded captions (e.g., ``a couple dancing on a Parisian rooftop at sunset''). Its hybrid encoder fuses vision tokens with text prompts, enabling effective \textit{cross-modal grounding} while mitigating hallucinations often seen in video encoders. To bridge the audio gap, we propose a two-stage approach:  
1. \textit{Audio Transcription}: We extract speech and texture using Whisper~\cite{radford2023robust}, selected for its robustness to background noise.  
2. \textit{Audio Knowledge Fusion}: We feed the transcriptions and sound classifications (e.g. ``dramatic music'') into Qwen-Audio~\cite{chu2023qwen}, which generates intent-aware descriptions (e.g. ``upbeat soundtrack with a lighthearted tone''). This modular design enables \textit{decoupled audio and video analysis}, while compensating for the absence of a unified audio-visual MLLM.

\subsubsection{Recommendation models}
Our framework is model-agnostic. To demonstrate its versatility, we evaluate two widely used models in academia and industry: the two-towers model, originally proposed by Youtube~\cite{covington2016deep}, and the generative model SASRec~\cite{kang2018self}.
Both models consist of an item and a user encoder. In the two-towers model, the user embedding is computed as the average embedding of the videos watched by the user, following~\cite{ni2023content}. In SASRec, the user encoder is implemented as a decoder-only Transformer. Its output  is further processed through a Residual MLP before computing the loss,  following the design of PinnerFormer~\cite{pancha2022pinnerformer}. The item encoder projects embedding features through a standard 1-layer Residual MLP with ReLU activations.  We use the same hyperparameters as~\cite{ni2023content} and do not perform any hyperparameter tuning. We do not here consider solutions that finetune vision models, such as \cite{ni2023content}, due to their significantly high computational cost.

\subsubsection{Evaluation}
We evaluate the performance of our recommendation task using HitRate (HR) and Normalized Discounted Cumulative Gain (nDCG) @ K, with K set to 10 and 30.
In line with recent evaluation standards~\cite{meng2020exploring}, we adopt a global time-based split with a 1-day rolling window to closely mimic production settings. For each of the last seven days,  we perform daily evaluations. On day $k$, the model is trained on data up to day k-2, with validation on day k-1 to determine the optimal number of training epochs using early stopping (patience$=$5). Once the best epoch count is identified, we retrain the model using data up to day k-1 and evaluate it on day $k$.
Unless specified otherwise, all performance comparisons (e.g., between Model A and Model B) are based on a paired t-test.

\subsection{Results}

Prompting an off‑the‑shelf MLLM to summarise each clip into natural‑language descriptions consistently improves performance across all tested modalities. On MicroLens‑100K, replacing raw background audio features with MLLM‑generated text boosts the two‑towers’ HR@10 from 0.0253 to 0.0405 and nDCG@10 from 0.0130 to 0.0214, a $\sim60\%$ relative gain (see Table~\ref{tab:main_results}). SASRec demonstrates similarly strong improvements, with a relative gain of $35\%$.
Converting audio waveforms to text recovers aspects such as theme, mood, and world knowledge, information typically lost in traditionally spectrogram-based representations. Similarly, adding scene‑level captions boosts HR@10 from 0.0393 to 0.0489 in the two‑towers model ($+24\%$) and lifts SASRec to 0.0482, outperforming raw frames and creator-written titles by $4\%$ to $18\%$. In essence, pixels show \emph{what happens on-screen}, titles reflect \emph{what the uploader hopes} will attract clicks, but MLLM-generated text captures \emph{why viewers might care}: conveying tone, parody, and cultural cues that standard recommenders can now begin to utilize. Because MLLM-generated descriptions condenses minute-long sequences into a single, coherent summary, they also help overcome the long‑horizon limitations of conventional vision-based features~\cite{wu2022memvit, vahdani2022deep}.

\textbf{Scaling encoders and backbones.} Table~\ref{table:size} explores two scaling strategies  while keeping the rest of the pipeline unchanged: (i) swapping the text encoder with a larger, higher-performing model  (Qwen embeddings 0.6B~\cite{zhang2025qwen3}, released in June 2025), and (ii) replacing the MLLM backbone with a larger variant (Qwen-VL 7B). While qualitative inspection reveals that Qwen-VL 7B generates richer captions with more grounded world knowledge, these upgrades do no yield  measurable improvements. This suggests that, for this dataset, the baseline MLLM already captures the most relevant information, and that merely increasing model size offers diminishing returns once a coherent video-level description is achieved. 

\begin{table}[t]
\small
\centering
\caption{\textbf{Performance of various modalities features on two widely used representative models: the two-towers and SASRec. The results show that MLLM-generated descriptions improve the performance in these recommendation systems.}}
\label{tab:main_results}
\setlength{\tabcolsep}{3.0pt}
\begin{tabular}{@{}lrrrr@{}}
\toprule
\makecell{\textbf{Representation}}      & \makecell{\textbf{HR@10}} & \makecell{\textbf{HR@30}} & \makecell{\textbf{nDCG@10}} & \makecell{\textbf{nDCG@30}} \\ 

\midrule
\multicolumn{5}{c}{\textit{Two-Towers (Youtube)~\cite{covington2016deep}}} \\
\midrule

Metadata &  0.0414 & 0.0721 & 0.0214 & 0.0286\\ 
Audio &   0.0253 & 0.0439 & 0.0130 & 0.0173 \\ 
Video &   0.0393 & 0.0729 & 0.0201 & 0.0280  \\ 
Metadata + video &   0.0428 & 0.0775 & 0.0222 & 0.0304   \\ 
MLLM audio & 0.0405 & 0.0742 & 0.0214 & 0.0294    \\ 
MLLM video &  \textbf{0.0489} & \textbf{0.0879} & \textbf{0.0264} & \textbf{0.355} \\

\midrule
\multicolumn{5}{c}{\textit{SASRec~\cite{kang2018self} (Generative recommendations)}} \\
\midrule
Metadata & 0.0460 & 0.0791 & 0.0249 & 0.0326  \\ 
Audio &  0.0338 & 0.0555 & 0.0186 & 0.0237  \\ 
Video &    0.0456    &  0.0829  & 0.0245 & 0.0332  \\
Metadata + video &    0.0462    &  0.0813  & 0.0246 & 0.0329  \\ 
MLLM audio & 0.0454 & 0.0816 & 0.0245 & 0.0330 \\ 
MLLM video &  \textbf{0.0482} & \textbf{0.0877}  & \textbf{0.0261} & \textbf{0.0353} \\ 
\bottomrule
\end{tabular}
\end{table}


\begin{table}[t]
\small
\centering
\caption{\textbf{Effect of encoder variants on SASRec performance.}}
\label{table:size}
\begin{tabular}{@{}lrrrr@{}}
\toprule
\textbf{Model}      & \makecell{\textbf{HR@10}} & \makecell{\textbf{HR@30}} & \makecell{\textbf{nDCG@10}} & \makecell{\textbf{nDCG@30}} \\ 
\midrule

Baseline & 0.0482 & 0.0877  & 0.0261 & 0.0353 \\ 

\midrule
\multicolumn{5}{c}{\textit{Larger text encoder}} \\
\midrule 
Qwen emb.~\cite{zhang2025qwen3} & 0.0479 & 0.0878  & 0.0256 & 0.0349 \\  

\midrule
\multicolumn{5}{c}{\textit{Larger MLLM}} \\
\midrule
Qwen-VL 7B & 0.0480 & 0.0862 & 0.0256 & 0.0346 \\ 
\bottomrule
\end{tabular}
\end{table}

\section{Conclusion}
We introduce a zero‑finetuning plug‑and‑play framework that transforms  raw audio‑video inputs into rich text descriptions using off‑the‑shelf MLLMs. These features can be seamlessly integrated into standard collaborative, content‑based, and generative recommenders. Without any additional training, our framework achieves up to 60\% relative gains on the MicroLens-100K dataset. As emerging omni-MLLMs~\cite{zhu2025extending}, such as Qwen Omni~\cite{xu2025qwen2} and Vita~\cite{fu2025vita}, continue to advance in jointly modeling vision, audio, and text, our approach  provides a low-barrier pathway for both research prototypes and production systems to improve recommendations by recognizing not only \emph{what} is on-screen, but also understanding \emph{why} it may resonate with individual viewers.


\bibliographystyle{ACM-Reference-Format}
\bibliography{sample-base}


\end{document}